\documentclass[lettersize,journal]{IEEEtran}
\IEEEoverridecommandlockouts
\usepackage{float}
\usepackage{threeparttable}
\usepackage{multirow}
\usepackage{makecell}
\usepackage{cite}
\usepackage{amsmath,amsfonts}
\usepackage{array}
\usepackage{booktabs} 
\usepackage[caption=false,font=normalsize,labelfont=sf,textfont=sf]{subfig}
\usepackage{textcomp}
\usepackage{stfloats}
\usepackage{url}
\usepackage{verbatim}
\usepackage{graphicx}
\usepackage{algpseudocode}
\usepackage{caption}
\usepackage{amsmath}
\usepackage[ruled,vlined]{algorithm2e}
\usepackage{algpseudocode} 
\usepackage{caption}
\usepackage{xcolor}
\usepackage{float}
\usepackage{makecell}

\usepackage{tabularx}
\usepackage{colortbl}
\definecolor{mygray}{gray}{.9}
\definecolor{mygray1}{gray}{.8}
\definecolor{mygray2}{gray}{.7}
\definecolor{mygray3}{gray}{.6}
\usepackage{diagbox}
\usepackage{enumitem}
\usepackage{tikz}

\usepackage[switch]{lineno,}
\usepackage{xcolor}

\newcommand*{\circledd}[1]{\lower.7ex\hbox{\tikz\draw (0pt, 0pt)
   circle (.4em) node {\makebox[0.25em][c]{\small#1}};}}

\newcommand*\circled[1]{\tikz[baseline=(char.base)]{
       \node[shape=circle,fill,inner sep=1pt] (char) {\textcolor{white}{\small#1}};}}

\usepackage{xcolor}
\definecolor{mygray}{gray}{.9}

\begin{document}

\title{PIMfused: Near-Bank DRAM-PIM with Fused-layer Dataflow for CNN Data Transfer Optimization}
\author{

Simei Yang, Xinyu Shi, Lu Zhao, Yunyu Ling, Quanjun Wang and Francky Catthoor~\IEEEmembership{}

}



\maketitle

\begin{abstract}


Near-bank Processing-in-Memory (PIM) architectures integrate processing cores (PIMcores) close to DRAM banks to mitigate the high cost of off-chip memory accesses. When accelerating convolutional neural network (CNN) on DRAM-PIM, performance is often constrained by cross-bank (or cross-PIMcore) data transfers, which are induced by the conventional layer-by-layer dataflow that enforces inter-bank (or inter-PIMcore) dependencies across successive CNN layers. To address this challenge, we propose \textit{PIMfused}, a hardware–software co-design that enables fused-layer dataflow for end-to-end CNN execution in near-bank DRAM-PIM. By adopting fused-layer dataflow, PIMfused improves data reuse and, more importantly, breaks inter-bank data dependencies, thereby optimizing cross-bank data transfers without sacrificing bank-level parallelism. We study the impact of buffer sizes and PIMcore parallelism (1-bank vs. 4-bank) on PIMfused using end-to-end ResNet18. We present three key takeaways and show that with 4-bank PIMcores, PIMfused achieves overall PPA gains over a GDDR6-AiM–like baseline, cutting memory cycles to 30.6\%, energy to 83.4\%, and area to 76.5\%.




\end{abstract}


\section{Introduction}
\label{Sec:Intro}

DRAM Processing-In-Memory (PIM) reduces the cost of off-chip memory accesses by placing processing cores (PIMcores) near DRAM banks, as in UPMEM~\cite{devaux2019true}, Samsung Function-in-Memory (FIM)~\cite{kwon202125} and SK Hynix AiM~\cite{he2020newton,kwon20221ynm}. These near-bank DRAM-PIM designs shorten data movement, lower data access latency, and exploit bank-level parallelism, making them effective for memory-intensive workloads such as large language models~\cite{park2024attacc}. Deep convolutional neural networks (CNNs), despite high compute demands, still suffer from frequent off-chip accesses due to limited on-chip memory sizes, which makes them common benchmarks in industrial PIM prototypes~\cite{he2020newton,kwon20221ynm,kwon202125,devaux2019true}. In this work, we focus on accelerating CNN workloads on near-bank DRAM-PIM architectures.


\begin{figure}[h]
	\centering
\includegraphics[width= 1\textwidth]{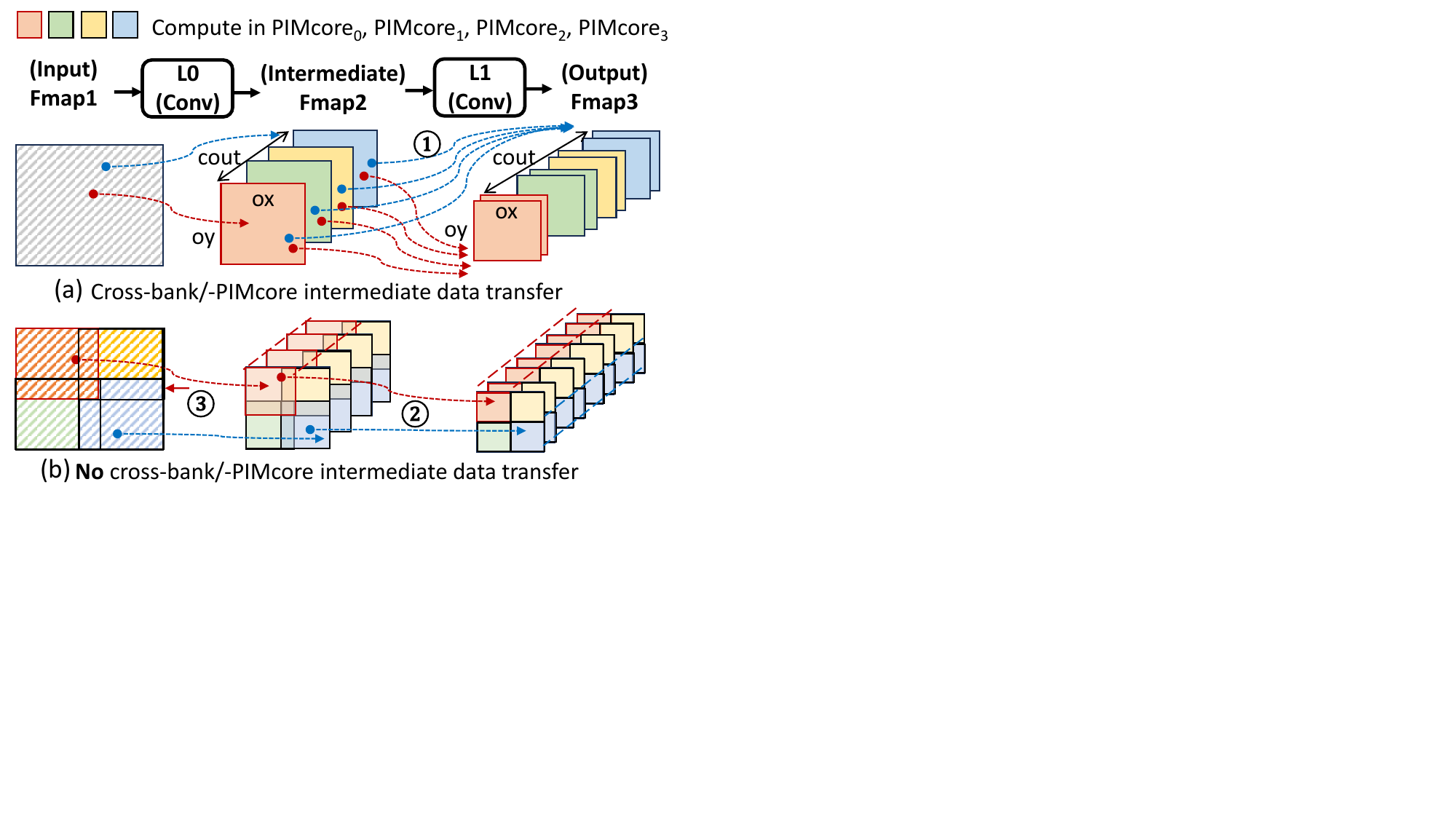}
	\vspace{-4.5cm}
	\caption{Dataflow comparison highlighting data transfers for PIMcore$_0$ and PIMcore$_3$: (a) Layer-by-layer dataflow. (b) Fused-layer dataflow. Fmaps (feature maps). }
	\label{fig:L2L_vs_Fused}
	\vspace{-0.5cm}
\end{figure}

When accelerating CNNs, current near-bank DRAM-PIM systems have to handle frequent cross-bank (or cross-PIMcore) data transfers\cite{devaux2019true}. A primary contributor to these transfers is the dataflow. Existing systems typically employ a layer-by-layer dataflow, in which CNN layers are processed sequentially\cite{kwon202125,he2020newton,kwon20221ynm}. Because each CNN layer depends on the outputs of the previous one(s), intermediate feature maps often need to be reorganized and redistributed across banks before the next layer starts. Fig.\ref{fig:L2L_vs_Fused}(a) illustrates \textit{cross-bank data transfer} in a layer-by-layer dataflow for two consecutive convolution layers (CONV, e.g., L0 and L1) on a DRAM-PIM with 4 banks and 4 PIMcores. Typically, each PIMcore computes partial output channels ($cout$) as in~\cite{kwon20221ynm,shin2023pimflow}. However, due to CNN computation rules, L1 requires all output channels from L0 (i.e., input channels for L1), forcing cross-bank transfers (e.g., \circledd{\small{\bf{1}}} highlight for PIMcore$_0$ and PIMcore$_3$). These transfers are expensive because banks (or PIMcores) typically cannot exchange data directly. Instead, data must be read from the source bank, sent over a shared internal bus, and routed, e.g.,via the host~\cite{kwon202125} or a channel-level Global Buffer (GBUF)~\cite{kwon20221ynm}, before being written to the target bank.

To optimize cross-bank transfer overhead, we explore a fused-layer dataflow that merges multiple consecutive layers (not only CONV followed by element-wise operations, but also multiple CONV layers) into a single computational kernel~\cite{gilbert2024looptree}. In systolic-array CNN accelerators, such fused-layer dataflow is widely used to exploit inter-layer data reuse~\cite{gilbert2024looptree}. Here, we demonstrate its potential for optimizing cross-bank data transfer in near-bank DRAM-PIM. Fig.~\ref{fig:L2L_vs_Fused}(b) shows a motivating example where fusing L0 and L1 allows decomposition into smaller, independent sub-kernels along the feature map’s spatial dimensions ($ox$, $oy$). Each PIMcore processes one sub-kernel across all output channels, reusing inter-layer intermediate data (e.g, via its near bank or local buffer within PIMcore) and eliminating cross-bank transfers (see \circledd{\small{\bf{2}}}). This preserves bank-level parallelism by breaking inter-bank (or inter-PIMcore) data dependencies, at the expense of some data duplication and redundant computation within fused layers (see \circledd{\small{\bf{3}}}). However, such costs remain small compared to the significant performance benefits of an optimized fused-layer dataflow~\cite{gilbert2024looptree}. For example, fusing the first 8 layers of ResNet18 into 4 tiles increases data replication by 18.2\% and redundant computation by 17.3\%, while delivering a 91.2\% performance improvement (see Section~\ref{sec:Exp3}).





In this paper, we propose \textbf{\textit{PIMfused}}, a hardware–software co-design for near-bank DRAM-PIM systems that leverages fused dataflow for end-to-end CNN execution. To the best of our knowledge, this is the first work to exploit fused-layer dataflow in near-bank DRAM-PIM for the optimization of cross-bank data transfers. Moreover, current near-bank PIMcore designs provide small buffer sizes due to DRAM die area constraints. For instance, AiM provides 2 KB of GBUF per channel~\cite{he2020newton,kwon20221ynm}, and FIM includes limited register files within each PIMcore~\cite{kwon202125}. These small buffers restrict data reuse and diminish the benefits of fused execution. To address this, we further analyze how different buffer sizes affect the performance, power, and area (PPA) of \textit{PIMfused}. We summarize our main \textbf{contributions} as follows.





\begin{itemize}
    \item  We propose the \textit{PIMfused} architecture, based on GDDR6-AiM~\cite{kwon20221ynm}, to enable end-to-end CNN execution while reducing data transfers through fused-layer dataflow. We also introduce custom PIM commands to coordinate computation and data transfer.

    \item We introduce the \textit{PIMfused} dataflow, which employs a fused-layer strategy for shallow CNN layers to decouple data dependencies across banks/PIMcores, and switches to a layer-by-layer strategy for deeper layers due to constraints imposed by CNN layer dimensions.

    \item We develop a PPA profiling framework to evaluate \textit{PIMfused} on end-to-end ResNet18. Our experimental study of buffer size and PIMcore parallelism provides three takeaways, showing that with 4-bank PIMcores, \textit{PIMfused} can outperform the baseline by cutting memory cycles to 30.6\%, energy to 83.4\%, and area to 76.5\%.
    
    
    


\end{itemize}

The remainder of this paper is organized as follows. Section~\ref{relatedwork} reviews related work and highlights our innovations. 
Section~\ref{Sec:Arch} and Section~\ref{Sec:Dataflow} introduced our proposed \textit{PIMfused} architecture and dataflow, respectively. Section~\ref{Sec:ExpTop} provides system PPA evaluations with different buffer size configurations. Finally, Section~\ref{sec:conclusion} concludes the paper.

 

\section{Related Work} 
\label{relatedwork}

This section surveys prior near-bank DRAM-PIM systems, focusing on their PIMcore design, cross-bank transfer mechanisms, and dataflow strategies for CNN execution.

UPMEM~\cite{devaux2019true} is widely recognized as the first commercialized near-bank PIM architecture. It integrates a general-purpose RISC core adjacent to each DDR4 bank, while relying on the host processor to manage cross-bank (or cross-PIMcore) data transfers. Samsung’s FIM architectures~\cite{kwon202125,medina2024bank} adopt a different design, employing a single PIMcore shared by two adjacent HBM banks within the same pseudo-channel. Each PIMcore incorporates a SIMD (Single Instruction, Multiple Data) unit with ALU (Arithmetic Logic Unit) capabilities and a set of local registers for executing micro-operations. Building on this design, FIM-DSE~\cite{medina2024bank} investigates logical register configurations ranging from 4 to 32 per PIMcore and evaluates performance across multiple DRAM standards, including HBM2, DDR4, GDDR5, and LPDDR4. SideDRAM~\cite{SideSIMD2025} integrates software-defined SIMD datapaths within a PIMcore and interfaces with row-wise wide registers to access multiple banks at once, achieving high memory access parallelism. The works of~\cite{kwon202125,medina2024bank,SideSIMD2025}
evaluate performance using kernels such as GEMV and CONV, with workloads averagely distributed across PIMcores. Full CNN execution requires reordering intermediate feature maps, which necessitates cross-bank data transfers handled by the host processor, similar to UPMEM.




McDRAMv2~\cite{cho2020mcdram} integrates seven compute PIMcores (16×16 systolic arrays) and one special core within each LPDDR4 channel. The special core handles cross-bank (or cross-PIMcore) data transfers and functions such as scaling, biasing, nonlinear operations, and (de)quantization. McDRAMv2 executes CNNs in a layer-by-layer manner, with intermediate data exchanged through the special core and channel-level shared buffers. While this design enables flexible cross-bank data transfers, it also increases control complexity and area overhead, and may become a performance bottleneck since all inter-PIMcore data must traverse the special core.

AiM architectures, such as Newton~\cite{he2020newton} and GDDR6-AiM~\cite{kwon20221ynm}, embed Multiply–Accumulate (MAC) units along with batch normalization and RELU support within each PIMcore, and use channel-level global buffers (GBUFs) to broadcast data to PIMcores and support cross-bank data transfers. PIMFlow~\cite{shin2023pimflow} extends the AiM architecture by incorporating 4 GBUFs to improve data reuse and provides compiler and runtime optimization. Compared to McDRAMv2~\cite{cho2020mcdram}, which uses a dedicated special core for cross-bank data transfers, AiM architectures with GBUFs offer a simpler hardware design. However, current AiM designs restrict the GBUF to reading one bank at a time sequentially, which prevents conflicts but limits parallelism. Optimizing cross-bank transfers (via GBUF and bank transfer) is therefore key to improving system performance.







Overall, the works discussed above~\cite{devaux2019true,kwon202125,medina2024bank,cho2020mcdram,he2020newton,kwon20221ynm,shin2023pimflow} adopt a layer-by-layer dataflow for CNN execution on near-bank DRAM-PIM systems. As previously illustrated in Fig.\ref{fig:L2L_vs_Fused}(a), the layer-by-layer dataflow requires reorganizing inter-layer intermediate data due to inter-bank (or inter-PIMcore) data dependencies. Such reorganization increases cross-bank data transfers, which are typically managed via the host\cite{kwon202125,medina2024bank}, a dedicated data-transfer core~\cite{cho2020mcdram}, or GBUF~\cite{he2020newton,kwon20221ynm,shin2023pimflow}. Compared to existing works, our work introduces the following \textbf{innovations}: (1) We optimize the AiM architecture~\cite{he2020newton,kwon20221ynm} to enable end-to-end CNN execution on near-bank DRAM-PIM by leveraging GBUF-based data broadcasting and cross-bank data transfers. In contrast, existing AiM designs focus primarily on MAC operations within PIM. (2) Instead of the conventional layer-by-layer dataflow, we exploit a fused-layer dataflow, previously unexplored in near-bank DRAM-PIM, that breaks dependencies across PIMcores and reduces cross-bank data transfers.
(3) While prior works explore different buffer configurations within PIM (e.g., 4–32 registers per PIMcore~\cite{medina2024bank}, 4 GBUFs at the channel level~\cite{shin2023pimflow}), We study the impact of buffer size on system performance for both channel-level GBUFs and local buffers (LBUFs) within each PIMcore.

\section{Proposed  Architecture \& PIM Commands}
\label{Sec:Arch}

This section presents \textit{PIMFused}, a near-bank DRAM-PIM architecture that exploits fused-layer dataflow to optimize cross-bank transfers and introduces custom PIM commands.




\subsection{PIMfused Architecture}
\label{Sec:Architecture}
\begin{figure}[h]
    \vspace{-0.4cm}
	\centering
\includegraphics[width= 0.78\textwidth]{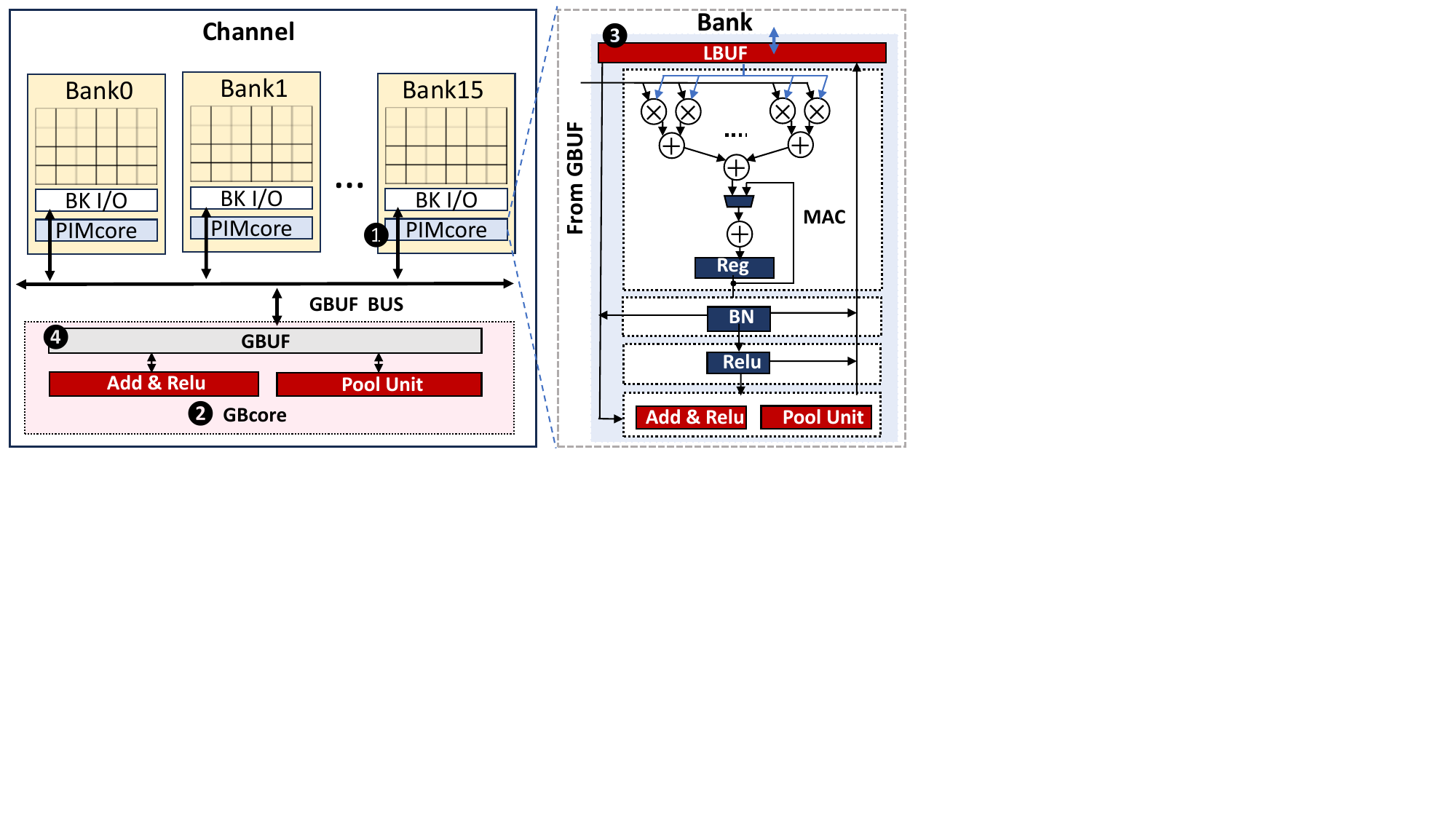}
	\vspace{-4cm}
	\caption{The \textit{PIMFused} architecture within a memory channel. }
	\label{fig:FusedArch}
	\vspace{-0.3cm}
\end{figure}


We propose the \textit{PIMFused} architecture, based on GDDR6-AiM~\cite{kwon20221ynm}, to leverage the channel-level GBUF for bank broadcasting and cross-bank data transfers. Fig.\ref{fig:FusedArch} illustrates the \textit{PIMFused} architecture, with new components compared to GDDR6-AiM\cite{kwon20221ynm} highlighted in red. The design features two types of processing cores. First, bank-level PIMcores (\circled{\small{\bf{1}}}) for fused-kernel operations, in which the functionality of each core depends on the requirements of the fused kernel. Compared to GDDR6-AiM, our PIMcores not only support MAC, Batch Normalization (BN), and RELU functions, but also include residual addition and pooling to enable more flexible kernel fusion. Second, a channel-level GBcore (\circled{\small{\bf{2}}}) for pooling, residual addition, and other data reduction tasks. By supporting residual addition and pooling in both PIMcores and the GBcore, the \textit{PIMFused} architecture provides greater flexibility for executing end-to-end CNNs. In addtion, to improve data reuse, we add a local buffer (LBUF, \circled{\small{\bf{3}}}) to each PIMcore alongside the original global buffer (GBUF, \circled{\small{\bf{4}}}). Given the complexity of PIMcore designs and the impact of buffer sizes, we evaluate two configurations to study system PPA trade-offs (Section~\ref{Sec:ExpTop}): (1) \textit{1-bank PIMcore}, where each bank has a dedicated PIMcore; (2) \textit{4-bank PIMcore}, where a single PIMcore is shared among four banks.






\subsection{Custom Command for PIM Execution}
\label{Sec:PIMcmd}

TABLE~\ref{table:CMDTable} lists the our custom PIM commands (CMD). For PIM computation, we introduce two CMDs, PIMcore\_CMP and GBcore\_CMP, for PIMcore and the GBcore, respectively. Each CMD supports different execution flags depending on its active components. In particular, PIMcore\_CMP provides four execution flags. For example, if a fused kernel includes CONV, BN, RELU, and POOL operations, two flags should be enabled: CONV\_BN\_RELU and POOL. A single PIMcore\_CMP command from the memory controller activates all PIMcores concurrently, enabling parallel execution.

\begin{table}[t]
\caption{Custom CMD for PIMFused Execution}
\renewcommand\arraystretch{1.5}
\scriptsize
\begin{center}
\begin{threeparttable}
\vspace{-0.3cm}

\begin{tabular}{|>{\raggedright\arraybackslash}m{1.2cm}<{\centering}|>{\raggedright\arraybackslash}m{1.9cm}<{\centering}|>{\raggedright\arraybackslash}m{4.4cm}
<{\centering}|}
\hline

\bf Type & \textbf{CMD} &  {\textbf{Description}}   \cr\hline

PIM &PIMcore\_CMP& Perform fused operations in all PIMcores\cr\cline{2-3}

Computation&GBcore\_CMP& Perform operations in GBcore\cr\hline


 &PIM\_BK2LBUF&\multirow{2}{*}{\shortstack[l]{Data transfer between all banks and LBUFs}}\cr\cline{2-2}
Data&PIM\_LBUF2BK&\cr\cline{2-3} 
Movement&PIM\_BK2GBUF&\multirow{2}{*}{\shortstack[l]{Data transfer between one bank and GBUF}}\cr\cline{2-2}
&PIM\_GBUF2BK&\cr\hline


\end{tabular}
    \begin{tablenotes}
    \footnotesize
    \item *Note: PIMcore execution flags: CONV\_BN, CONV\_BN\_RELU, POOL, ADD\_RELU; GBcore execution flags: POOL, ADD\_RELU; 
   \end{tablenotes}
    \end{threeparttable}
    \vspace{-0.7cm}
\label{table:CMDTable}
\end{center}
\end{table}


For data movement, PIM\_BK2LBUF and PIM\_LBUF2BK handle data transfers between banks and LBUFs, allowing multiple PIMcores to access their local banks concurrently. In contrast, PIM\_BK2GBUF and PIM\_GBUF2BK manage data transfers between banks and the channel-level GBUF. Following the AiM architecture~\cite{he2020newton,kwon20221ynm}, these GBUF-related CMDs require sequential transfers to simplify control and avoid data conflicts. The memory controller fetches data from one bank at a time and writes it to the GBUF before proceeding to the next bank. As a result, transfers to the LBUF are faster due to the high parallel bandwidth available to all PIMcores (or LBUFs), while transfers to the GBUF are slower because of sequential access. In \textit{PIMfused}, data cannot move directly from an LBUF (or PIMcore) to the channel-level GBUF. Instead, all transfers between the LBUF and GBUF go through the memory banks, ensuring predictable behavior and simplified control, as in AiM~\cite{he2020newton,kwon20221ynm}.




\section{Proposed Dataflow}
\label{Sec:Dataflow}

\begin{figure*}[t]
	\centering
\includegraphics[width= 1.09\textwidth]{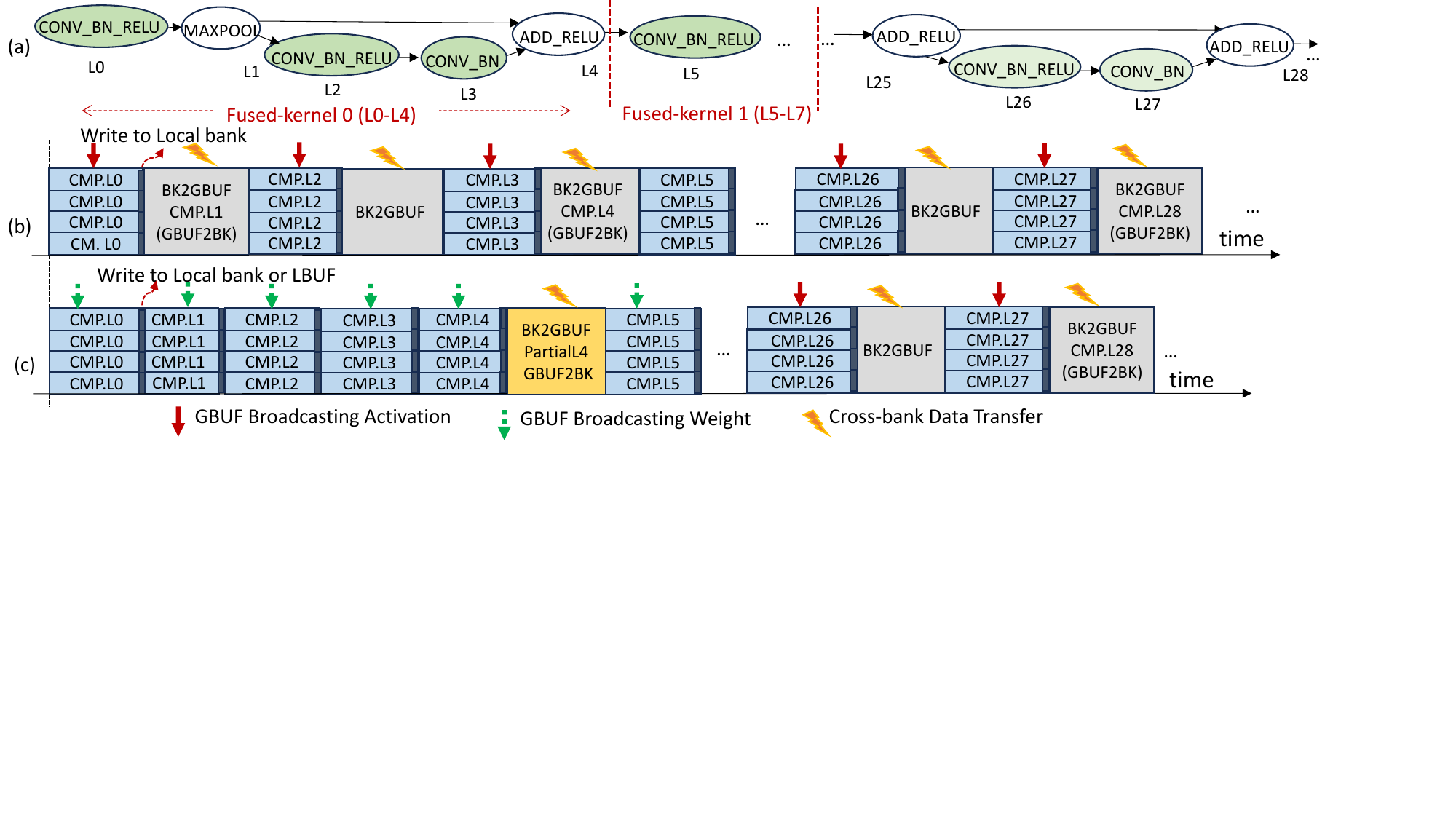}
	\vspace{-5.6cm}
	\caption{(a) CNN graph example. (b) Layer-by-layer dataflow on PIMfused with POOL and ADD\_RELU executed on GBcore. (c) PIMfused dataflow, storing intermediate data in local bank or LBUF in PIMcore.  (*BK2GBUF: data transfer from bank to GBUF; *GBUF2BKF: data transfer from GBUF to bank.)}
	\label{fig:ExecutionTrance}
	\vspace{-0.3cm}
\end{figure*}

This section presents the proposed \textit{PIMfused} dataflow, designed to reduce cross-bank transfer overhead. It is a hybrid strategy that combines fused-layer and layer-by-layer dataflows for end-to-end CNN execution. The two dataflows differ in how computations are mapped to PIMcores and how data are reused and broadcast through the LBUF and GBUF.



\begin{itemize}
    \item \textit{Layer-by-layer dataflow}: 
    Each PIMcore computes a subset of output feature maps by tiling across output channels ($cout$), as shown in Fig.~\ref{fig:L2L_vs_Fused}(a).  We distribute the weights of each CONV layer across banks according to these $C_{out}$ segments. During execution, LBUFs locally reuse CNN weights, while the GBUF reuses activations and broadcasts them to all PIMcores~\cite{he2020newton,kwon20221ynm}.

    \item \textit{Fused-layer dataflow}: 
    Each PIMcore computes a subset of output feature maps by tiling across spatial dimensions ($ox$, $oy$), as shown in Fig.~\ref{fig:L2L_vs_Fused}(b). We partition input activations across banks based on these spatial segments ($ox$, $oy$) defined by the fused-layer strategy. During execution, LBUFs locally reuse CNN activations, while the GBUF reuses weights and broadcasts them to all PIMcores.
    
    
    

\end{itemize}



The \textbf{\textit{PIMfused dataflow}} exploits fused-layer processing for shallow layers of a CNN, which typically have large spatial dimensions ($ox$,$oy$), and switches to layer-by-layer processing for deeper layers, which have many output channels ($cout$). Fig.\ref{fig:ExecutionTrance} compares the proposed \textit{PIMfused} dataflow with the conventional layer-by-layer dataflow on a full CNN (e.g.,ResNet18) running on the \textit{PIMfused} architecture, comprising 4 PIMcores and 1 GBcore. Fig.\ref{fig:ExecutionTrance}(a) presents the CNN application graph. Throughout this paper, element-wise fusion operations (e.g., CONV\_BN\_RELU) are applied by default and treated as a single layer.

Fig.\ref{fig:ExecutionTrance}(b) illustrates the baseline layer-by-layer dataflow. PIMcores execute CONV layers (e.g., CONV\_BN\_RELU), while the GBcore handles non-CONV layers (e.g., POOL, ADD\_RELU). In the first layer (L0), each PIMcore (or LBUF) loads its weights (different $C_{out}$) from the banks, and the GBUF broadcasts input activations to all PIMcores\cite{he2020newton,kwon20221ynm}. Due to limited LBUF capacity, PIMcores write L0 outputs back to local banks. The GBUF then fetches these outputs (via PIM\_BK2GBUF CMD in Table~\ref{table:CMDTable}) for L1 computation in the GBcore. If the GBUF overflows, part of the data is written back to banks (via PIM\_GBUF2BK CMD in Table~\ref{table:CMDTable}). 
Afterward, the system broadcasts L1 outputs (i.e., L2 inputs) to PIMcores for L2 computation, and the intermediate results are written to banks again. This process repeats until the application completes. In this dataflow, the GBUF frequently transfers intermediate data between banks and itself, either for GBcore execution or for broadcasting to PIMcores, leading to high cross-bank transfer overhead.

Fig.\ref{fig:ExecutionTrance}(c) shows our proposed PIMfused dataflow, which applies fused-layer execution to shallow CNN layers and layer-by-layer execution to deeper layers. In our CNN example (Fig.\ref{fig:ExecutionTrance}(a)), layers L0–L4 form the first fused-kernel and L5–L7 form the second. For the first fused-kernel, all PIMcores (or LBUFs) fetch L0 inputs from banks, each handling a different spatial segment ($ox,oy$) (Fig.\ref{fig:L2L_vs_Fused}(b)), while the GBUF broadcasts weights. PIMcores write L0 outputs to the LBUF when capacity allows or to local banks. With limited LBUFs, intermediate data spills across LBUF and DRAM, whereas an ideal LBUF sets the upper performance bound at the cost of extra energy and area (see Section\ref{sec:Exp3}). After L0, PIMcores fetch intermediate data from banks or LBUFs to compute L1, repeating this process until the fused-kernel completes. At fused-kernel boundaries, the GBUF reorganizes intermediate data for the next fused-kernel (orange boxes in Fig.\ref{fig:ExecutionTrance}(c)), as boundary-layer inputs may span multiple banks or PIMcores. Once all fused-kernels finish, execution returns to the layer-by-layer dataflow for deeper layers. Compared to layer-by-layer execution (Fig.\ref{fig:ExecutionTrance}(b)), \textit{PIMfused} reduces cross-bank data transfers within each fused-kernel.

\section{Experimental Evaluation}
\label{Sec:ExpTop}

\subsection{Experimental Methodology}

\begin{figure}[h]
    \vspace{-0.5cm}
	\centering
\includegraphics[width= 1.17\textwidth]{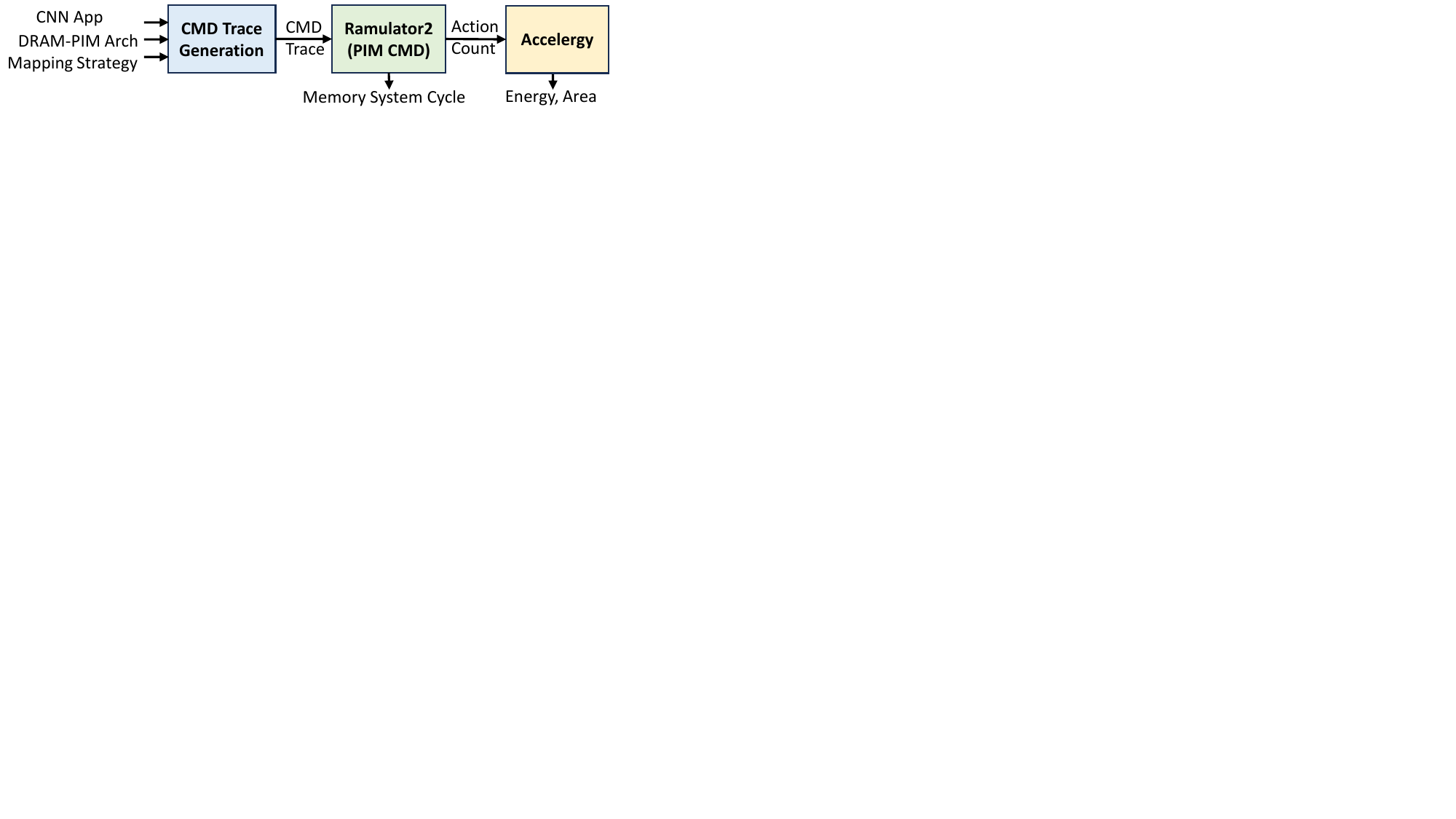}
	\vspace{-10.9cm}
	\caption{Overview of our profiling framework.}
	\label{fig:ProfilingFramework}
	\vspace{-0.2cm}
\end{figure}

\subsubsection{Simulation} We developed a profiling framework for fast evaluation of system PPA for \textit{PIMfused}. Fig.\ref{fig:ProfilingFramework} illustrates the framework, which is built on the open-source Ramulator2\cite{luo2023ramulator} and Accelergy~\cite{wu2019accelergy}. We extend Ramulator2~\cite{luo2023ramulator} to model the \textit{PIMfused} design on a 16-bank GDDR6 channel by adding custom PIM commands (see Table~\ref{table:CMDTable}). For each simulation, we generate command traces based on the target CNN, DRAM-PIM architecture (e.g., number of PIMcores, buffer sizes), and mapping strategy (e.g., layer-by-layer, fused-layer). Ramulator2 reports memory system cycles, which we use as the performance metric, following ~\cite{park2024attacc}.







We use Accelergy~\cite{wu2019accelergy} to estimate component-level area and energy. It models a system as a hierarchy of components, each with a plugin specifying area and energy. Combining these attributes with action counts, Accelergy computes total energy for a workload. For SRAM buffers, we use Accelergy’s built-in CACTI integration at 22nm. For PIMcore and GBcore, we build compound components from primitive units (e.g., adders, multipliers, dividers, comparators, barrel shifters) and characterize them with our in-house post-synthesis data at 22nm. For DRAM-PIM, we configure Accelergy’s abstract DRAM model, scale GDDR6 access energy from GDDR5 values, and assume near-bank accesses consume 40\% of this energy because they bypass I/O costs. We model the internal bus between banks and the GBUF with wire models.



\subsubsection{Benchmark} We use ResNet18 as the representative benchmark. Since the proposed \textit{PIMfused} employs a hybrid dataflow, we consider two workload scenarios. First, \textit{ResNet18\_First8Layers} executes the first 8 layers on the DRAM-PIM systems to quantify the benefits of pure fused-layer dataflow versus layer-by-layer dataflow. Second, \textit{ResNet18\_Full} executes the entire ResNet18 end-to-end.

\subsubsection{Baseline} 

As discussed in Section~\ref{Sec:Architecture},  we believe the most representative baseline to compare our proposed PIMfused system to is the architecture in the GDDR6-AiM~\cite{kwon20221ynm}. To enable a fair comparison with our end-to-end CNN support, we add a GBcore (see Fig.\ref{fig:FusedArch}) to form a GDDR6-AiM-like system. This system, with the default buffer configuration (GBUF=2KB, LBUF=0)~\cite{kwon20221ynm}, serves as our baseline. All experimental results are reported relative to this baseline.  We summarize the DRAM-PIM systems evaluated as follows:




\begin{itemize}

\item  \textbf{\textit{AiM-like}}: The baseline system includes a GBcore and 16 lightweight PIMcores (\textit{1-bank PIMcore}), each supporting \textit{MAC}, \textit{BN}, and \textit{RELU}, as in GDDR6-AiM~\cite{kwon20221ynm}. It uses a layer-by-layer dataflow, with MAC operations on PIMcores and non-MAC operations on the GBcore. 



\item \textbf{\textit{Fused16}}: The proposed \textit{PIMfused} system (Fig.~\ref{fig:FusedArch}) includes a GBcore and 16 PIMcores (\textit{1-bank PIMcore}). In end-to-end execution of ResNet18, each fused layer is divided into 4×4 tiles across the $ox$ and $oy$ dimensions. The first 8 layers form the first fused kernel, and the next 7 form the second. The remaining layers that cannot fit evenly into a 4×4 tiling follow a layer-by-layer dataflow.

\item \textbf{\textit{Fused4}}: The proposed \textit{PIMfused} system consists of a GBcore and 4 PIMcores (\textit{4-bank PIMcore}). In end-to-end execution of ResNet18, each fused layer is tiled into 2×2 segments across $ox$ and $oy$. The first 8 layers form the first fused-kernel, the next 7 layers form the second, and another 7 layers form the third. The remaining layers, which cannot fit evenly into a 2×2 tiling, execute using the layer-by-layer dataflow.

\end{itemize}

In our experiments, we aim to evaluate the benefits of \textit{PIMfused} across varying LBUF and GBUF sizes to study how buffer capacity affects PPA. We denote buffer configurations as G$m$K\_L$n$, where GBUF=$m$KB and LBUF=$n$B.






\subsection{Evaluation of Increasing GBUF Size Without LBUF}
\label{sec:Exp1}

\begin{figure}[h]
    \vspace{-0.2cm}
	\centering
\includegraphics[width= 1.1\textwidth]{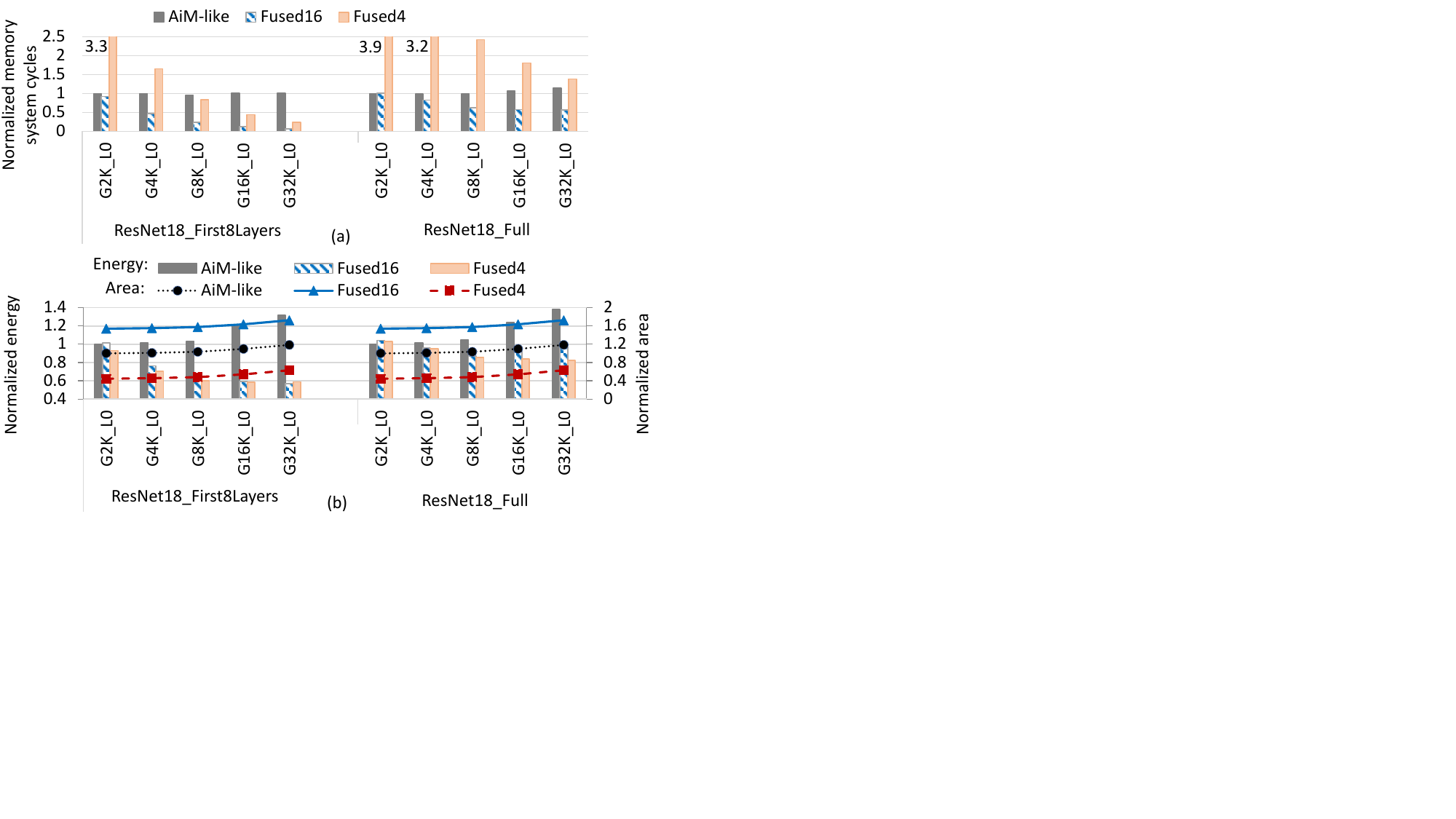}
	\vspace{-4.6cm}
	\caption{Normalized system PPA with increasing GBUF and no LBUF (w.r.t. AiM-like with G2K\_L0).}
        \vspace{-0.3cm}
	\label{fig:Exp1}
\end{figure}



Fig.~\ref{fig:Exp1} shows the normalized PPA of the three DRAM-PIM systems as GBUF increases with no LBUF, using a GDDR6-AiM-like setup\cite{kwon20221ynm} (AiM-like with G2K\_L0) as the baseline. We make four observations. First, AiM-like shows minimal performance improvement as GBUF grows, since 2KB already suffices to reuse activations. Second, Fused16 and Fused4 benefit from larger GBUF, which reduces memory cycles by enabling better weight reuse. Unlike CNN activations, which are mostly reused within a single layer, weights are larger and shared across layers, thus increasing GBUF reduces frequent cross-bank transfers. Third, with GBUF = 32KB, Fused16 reduces memory cycles to 6.5\% for \textit{ResNet18\_First8Layers} and 57.7\% for \textit{ResNet18\_Full}, at the cost of 55.1–72.4\% higher area. The smaller reduction in the full model occurs because deeper fused layers need more weight storage in GBUF, and fused-layer benefits are partially offset by layer-by-layer execution. Four, Fused4 underperforms AiM-like and Fused16 for \textit{ResNet18\_Full} due to lower PIMcore parallelism, but occupies only 44.6–63.1\% of the baseline area.

\textbf{\textit{Key Takeaway 1:} } \textit{
While GBUF=2KB is enough for layer-byl-ayer dataflow to reuse CNN activations, the PIMfused dataflow requires a larger GBUF (e.g., 8KB, 16KB) to support effective weight reuse.}


\subsection{Evaluation of Increasing LBUF Size With a Fixed GBUF}
\label{sec:Exp2}

\begin{figure}[H]
    \vspace{-0.3cm}
	\centering
\includegraphics[width= 1.1\textwidth]{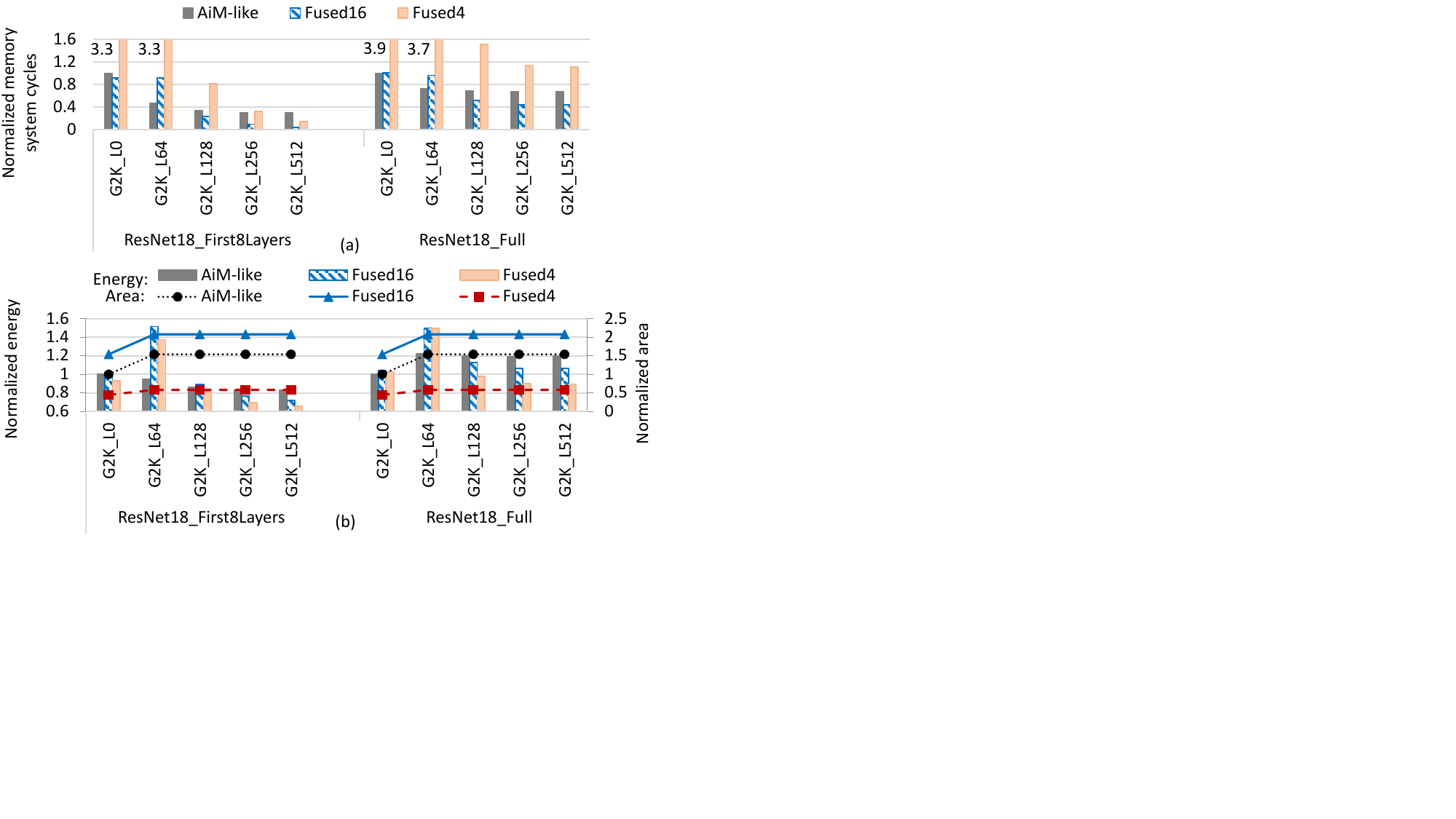}
	\vspace{-4.5cm}
	\caption{Normalized system PPA with increasing LBUF and fixed GBUF=2KB (w.r.t. AiM-like with G2K\_L0).}
        \vspace{-0.2cm}
	\label{fig:Exp2}
\end{figure}

Fig.~\ref{fig:Exp2} shows the normalized PPA of three DRAM-PIM systems as LBUF increases, with GBUF fixed at 2 KB. Overall, larger LBUF improves performance by enabling weight reuse in AiM-like and activation reuse in Fused16 and Fused4, with gains saturating after 256B. In \textit{ResNet18\_First8Layers}, adding a 64–512B LBUF reduces memory cycles to 30.2\%, 3.8\%, and 14.2\% of the baseline for AiM-like, Fused16, and Fused4. For \textit{ResNet18\_Full}, the reduction is smaller (67.9\%, 43.7\%, and 1.1×) because layer-by-layer execution in deeper layers diminish the benefits of fused-layer execution in shallow layers. In this case, Fused4 performs worse than Fused16 and AiM-like in memory cycles due to lower PIM-core parallelism, but uses only 44.6–58.1\% of the baseline area. Increasing LBUF from 64B to 512B adds little area overhead, since small SRAMs ($<$1KB) are dominated by peripheral circuitry in CACTI models.

\textbf{\textit{Key Takeaway 2:}} \textit{A small LBUF (e.g., 128B, 256B) per PIMcore can achieve high performance and energy efficiency in PIMfused by enabling reuse of CNN activations.}

\subsection{Evaluation of Increasing both GBUF and LBUF Sizes}
\label{sec:Exp3}
\begin{figure}[h]
    \vspace{-0.41cm}
	\centering
\includegraphics[width= 1.1\textwidth]{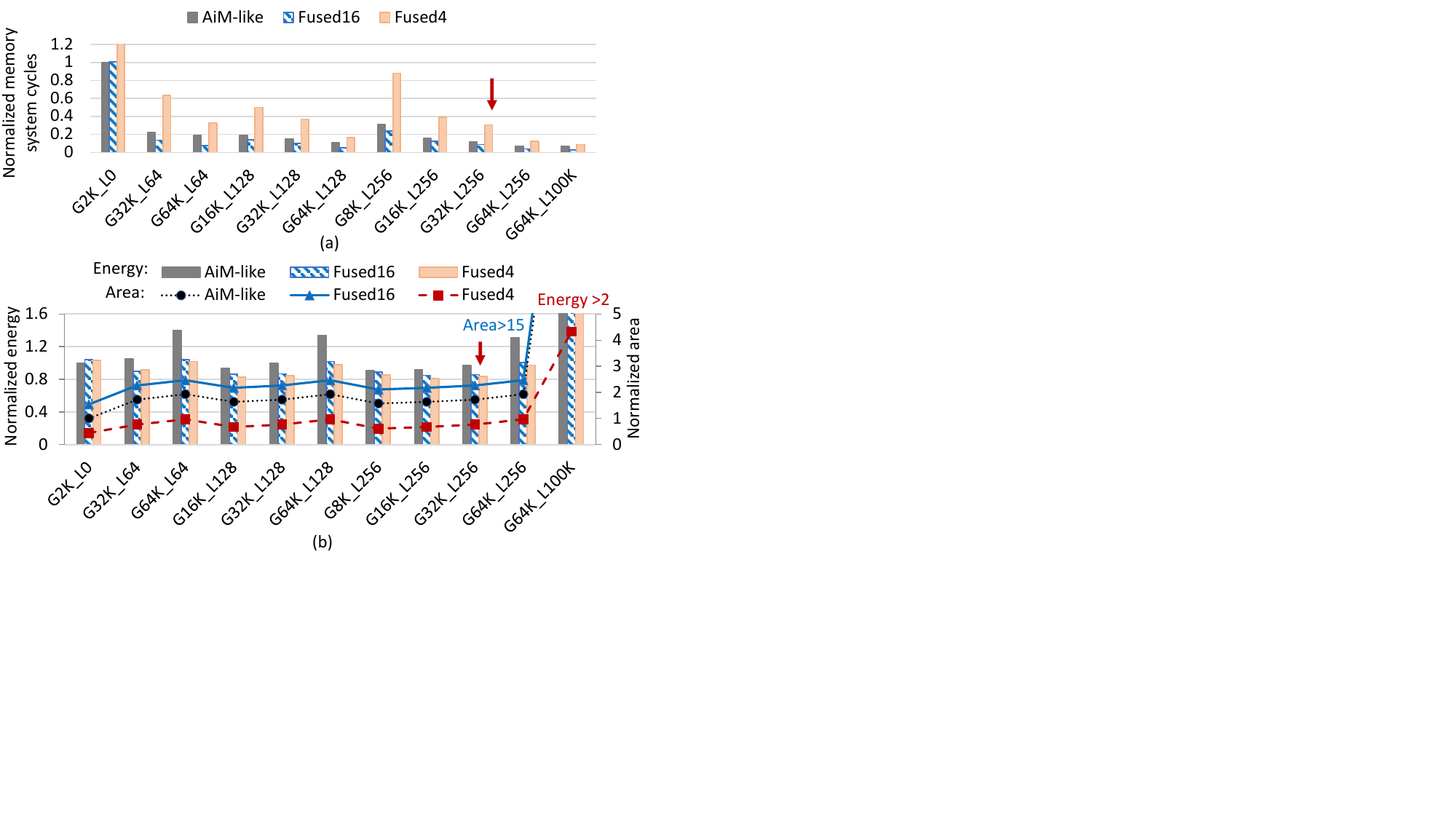}
	\vspace{-4.2cm}
	\caption{Normalized system PPA with increasing sizes of LBUF and GBUF (w.r.t. AiM-like with G2K\_L0), for \textit{ResNet18-Full}. }
        \vspace{-0.2cm}
	\label{fig:Exp3}
\end{figure}

Fig.\ref{fig:Exp3} shows the normalized PPA of the three DRAM-PIM systems for \textit{ResNet18\_Full} across varying LBUF and GBUF sizes. Compared to Sections\ref{sec:Exp1} and \ref{sec:Exp2}, increasing both buffers further improves performance and energy efficiency. With these configurations, Fused16 and Fused4 consume similar energy, with Fused4 slightly more efficient because fewer tiles (4 vs. 16) result in less data duplication and redundant computation. Fused16 achieves the highest performance at a higher area cost, while Fused4 offers the best area efficiency with slightly lower performance but still outperforms the AiM-like design. This demonstrates a Pareto trade-off between Fused4 and Fused16 in terms of performance versus area. Overall, our proposed \textit{PIMfused} with 4-bank PIMcores (Fused4) can outperform the AiM–like baseline across all PPA metrics. In particular, at G32K\_L256, Fused4 reduces memory cycles to 30.6\%, lowers energy to 83.4\%, and increases area to 76.5\% of the baseline. While G64K\_L256 further reduces memory cycles, it incurs higher energy and area costs. In an extremely large LBUF configuration (e.g., G64K\_L100K) that can hold all intermediate data of fused kernels, performance is similar to G64K\_L256, but energy and area rise dramatically.

\textbf{\textit{Key Takeaway 3:} } \textit{Exploring buffer sizes for both LBUFs and GBUF achieves better PPA than increasing only one. Extremely large LBUFs are unnecessary for near-optimal performance in PIM-fused systems.}

\section{Conclusion}
\label{sec:conclusion}

This paper proposes \textit{PIMfused}, a hardware–software co-design that exploits fused-layer dataflow in near-bank DRAM-PIM to reduce cross-bank data transfers during end-to-end CNN execution. The \textit{PIMfused} architecture integrates two types of processing cores (bank-level PIMcores and channel-level GBcore) and two types of buffers (LBUF in PIMcores and GBUF in GBcore). Its dataflow applies fused-layer execution to shallow CNN layers and switches to layer-by-layer execution in deeper layers. We  develop a PPA profiling framework to study the impact of buffer sizes and PIMcore parallelism. Experimental results on end-to-end ResNet18 reveal three key takeaways, showing that \textit{PIMfused} with 4-bank PIMcores delivers overall PPA improvements over a GDDR6-AiM–like baseline, reducing memory cycles to 30.6\%, energy to 83.4\%, and area to 76.5\%. In future work, we plan to evaluate additional benchmarks, particularly training workloads, and to extend our PPA profiling framework for improved DRAM energy estimation.









\bibliographystyle{IEEEtran}  
\bibliography{references}     %
\end{document}